# A co-Design approach to develop a smart cooking appliance.
# Applying a Domain Specific Language for a community supported appliance.

Matteo Zallio[1,*], Paula Kelly[2], Barry Cryan[3], Damon Berry[2]

[1]University of Cambridge, Department of Engineering, Trumpington st, Cambridge, UK
[2]Technological University Dublin, School of Electrical & Electronic Engineering, Ireland
[3]Saint John of God Community Services, Dublin, Ireland
mz461@cam.ac.uk; {paula.kelly, damon.berry}@tudublin.ie; barrycryan1111@gmail.com

**Keywords:** Co-Design approach, Assistive Technology, Domain Specific Language, Community Supported Appliance, Connected Home, Internet of Things, Activities of Daily Living, Cooking Instruction.

**Abstract**

Our environment, whether at work, in public spaces, or at home, is becoming more connected, adaptive and increasingly responsive. This is due in part to the significant number of sensors that detect information and data around us. Connected technologies are also emerging as a support to address daily challenges for people with disabilities and for ageing people, bringing care from hospitals into the community and increased provision of health services into homes. Meal preparation, even when it involves simply heating ready-made food, can be perceived as a complex process for people with disabilities. This research aimed to prototype, using a co-Design approach, a Community Supported Appliance (CSA) by developing a Domain Specific Language (DSL), precisely created for a semi/automated cooking process. The DSL was shaped and expressed in the users' idiom and allowed the CSA to support independence for users while performing daily cooking activities.

## 1. Introduction

The concept of an environment characterized by pervasive technologies, IoT-based de-vices, and shared data emerged more than two decades ago [1]. Recently, several companies in the consumer electronics market explored the concept of a connected kitchen, where people can benefit from the assistance of IoT devices during the cooking process [2]. Examples are the Counter Intelligence, from the Massachusetts Institute of Technology [3], the IBM Watson system [4] and Siemens Connected appliances with Home Connect [5]. The Ubiquitous Computing (UC) paradigm emphasizes the cooperation, interoperability, data sharing and interaction between different computational devices [6]. UC can also provide support for people with disabilities and ageing people, bringing care from hospitals into the community and helping to introduce e-health services into homes [7]. Connected homes and responsive appliances can support people to perform Activities of Daily Living (ADL) more independently with dignity [8]. One of the most challenging and important ADLs is to prepare a meal and eat [9]. During a given cooking process, people are required to be involved in different related activities and must engage in several subtasks. However, if such tasks are undertaken by a specific user who may suffer from physical and intellectual disabilities, the activity of cooking could become an intimidating process. This article, following previous research from the authors [10], reports on the development of a Community Supported Appliance (CSA), through the creation a Domain Specific Language (DSL), with a participatory co-Design approach. The CSA is composed of a responsive microwave oven that enables people with disabilities to prepare meals autonomously. According to Fowler, a DSL should be explained and expressed in the idiom of the users and accordingly the whole concept of the CSA was user-centric [11].

## 2. State of the art: Ubiquitous computing for the cooking process

The unification of different technologies and representations, especially between different devices, has always been a significant challenge in UC. Different works sought to re-strict the field of complexity and standardize how data and information are used. McDonald et al., developed HomeML to enable sharing of data sets from smart home environments [12], while Brock developed an attempt to create the Physical Markup Language (PML) as a common "language" for describing physical objects, processes and environments [13]. Later Krieg-Bruckner et al., researched how the complexity of a Formal Modelling for an application domain such as cooking can add value to the process of preparing recipes and cooking certain foods [14]. Furthermore, a compositional adaptation approach for cooking recipes represented as cooking workflows has been presented by Muller et al. [15]. Reichel et al., investigated and developed MAMPF, a multifunctional, adaptive meal preparation facility, as a versatile and adaptive kitchen system allowing inexperienced users to create recipes [16]. Russo et al., supported people's memory and coordination by using RFID tags for cooking instructions [17]. Miyawaki et al., developed a cooking support system for persons with brain dysfunction, through a multimodal interface of cooking-navigation to easily show instructions [18]. Tee et al., developed a Visually Enhanced Recipe Application, a language for communicating cooking information to systematically map the instructions from text to a primarily visual representation for persons with language impairments [19]. Ide et al., developed a method that detects difficult descriptions for an inexperienced user in an existing text recipe, and supplements them with multimedia content to facilitate the understanding of the recipe [20]. Most of the research addressed the issue of automating processes from different points of view. However, this work differentiates itself in four key aspects: (1) an accessible user interface with graphic representation of a DSL, (2) user empowerment from the community in managing and creating new sets of cooking instructions, (3) facilitation of a continued learning process for the user even when the microwave oven is in use and (4) the use of additional sensors to improve the user experience and safety features.

## 3. Research Methodology: designing in the idiom of the users
### 3.1 A participatory co-Design approach

The participatory co-Design approach enables a wide range of people to make a creative contribution to solve a problem [21]. Based on the authors' experience from a previous project [10], the co-Design approach was used to develop the structure, the grammar and the representation of the language for the cooking instructions. The research group conducted weekly meetings with service users and front-line care staff of Saint John of God Community of Dublin, an Irish-based social care service provider. As a result of several focus groups, each with 4 to 6 participants, it emerged that people with intellectual disabilities would be more likely to interact with a microwave oven if they had a graphical representation of the steps to guide them in meal preparation. A first draft of a simple DSL tailored according to the "idiom of the users" was then developed. Following this, archetypes [22], a graphical representation of the language, as well as short text information and audio reproductions of the indexed text were created (Fig. 1).

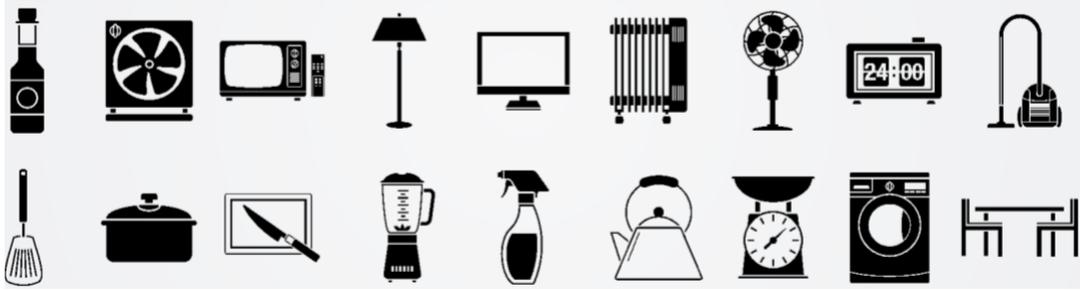

**Figure 1.** Sample icons used in the participatory co-Design approach.

User feedback was elicited around specific meal preparation needs according to an iterative process that involved users as the main designers of the graphical language. While the DSL is ultimately rendered in a JSON format, its representation follows users' preferences and can be designed using both a Unified Modeling Language (UML) class and a state diagram. The DSL of the cooking activities was represented first in UML to allow users to create cooking instructions as well as providing a user-friendly way of generating the DSL representation (in JSON format) (Fig. 2).

### 3.2 A simple multimodal language for cooking instructions

A UML class diagram [23] was created to capture the static grammar and terminology of the system. A UML state diagram was further created to represent the dynamic changes in states of the system in response to system events and a JSON data-interchange format was adopted to represent both the user and device instructions. In contrast to other works, this representation does not place emphasis on food ingredients or on the recipes, but on the interaction language and methodology that users suggested to adopt, to enable a community supported definition of cooking instructions.

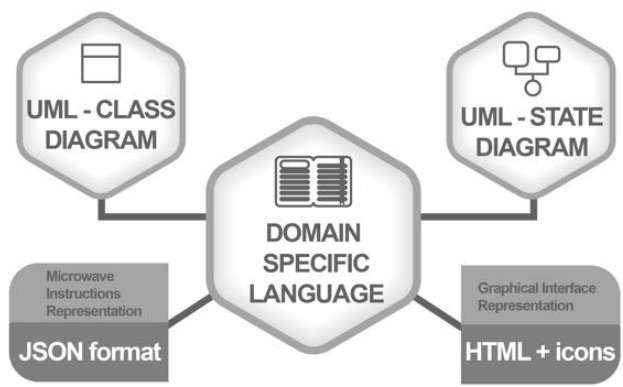

**Figure 2.** Representation of the general Domain Specific Language of the CSA.

In this way, the DSL facilitates community members to create and share interactive meal preparation instructions involving the microwave oven. The UML class diagram (Fig. 3) illustrates the static design for the DSL of the system and the following paragraphs describe some details about its main classes. The

"CookingInstructionSet" is a container class for a sequential set of cooking instructions. Multiple sets of instructions with varying levels of detail can be created for each individual food product, each set catering for a particular set of user abilities. The "Instruction" class represents a single instruction in a "CookingInstructionSet". It is a parent class of the "DeviceInstruction" and "UserInstruction" classes. The "DeviceInstruction" class represents the microwave oven instructions to heat a food product. Cooking time and cooking power will provide the parameter values to turn on the microwave for a specified period of time at a specified power setting respectively. The "UserInstruction" class focuses on providing multimedia instructions to the user that are comprised principally of still images and short audio clips. The "TransitionEvent" represents a transition event that signals the end of one instruction and the beginning of the next. Further "TransitionEvents" were introduced, including "DoorOpen", "DoorClosed", "WeightChange" and "SmokeDetected" events. The "Activation" class represents activation events such as on/off, true/false. Therefore, they will be associated with the elements in the "DeviceInstruction" class:

1. Light: Bool – this will result in the microwave light being on or off. It will be on for cooking and waiting periods and when the door is open.
2. Smoke Alarm: Bool – this will not result in the smoke alarm sounding if smoke has been detected. It will result in some message being displayed to the screen and a non-startling audio being played on the speakers.
3. Carousel: Bool – when the product is being heated and cooled the carousel will be turning. Although there is no safety issue associated with it, the carousel should not be turning if the microwave door is open.
4. Magnetron: Bool – the magnetron will be on when the product is being heated. The magnetron will never be on if the microwave door is open as it is a major safety issue.

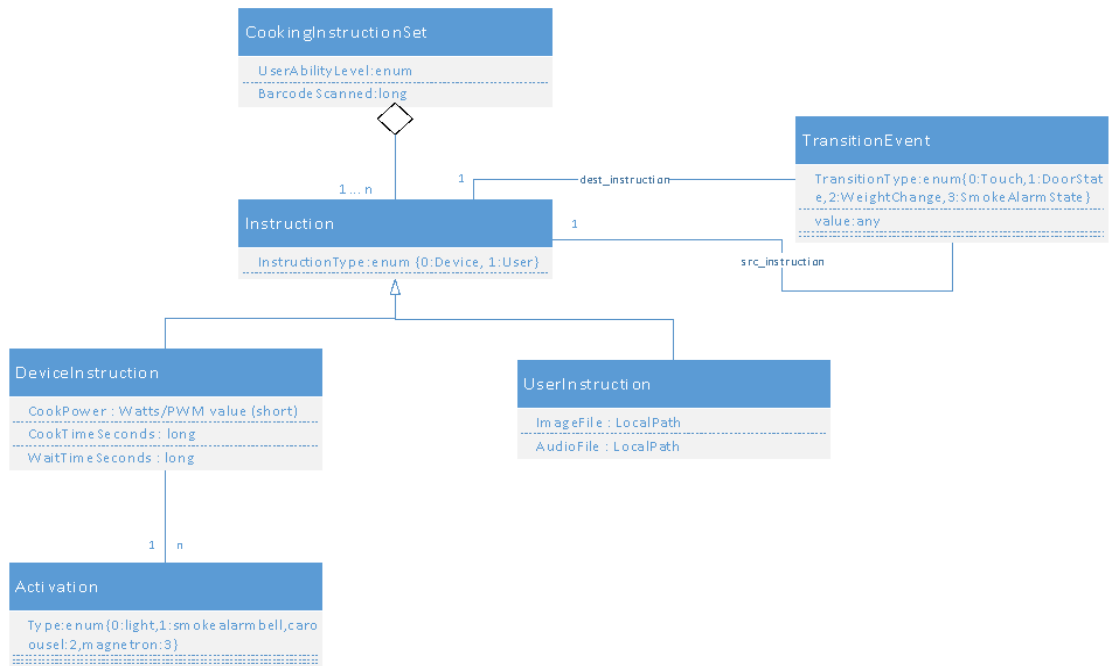

**Figure 3.** Representation of the UML class diagram for the CSA.

The following UML state diagram (Fig. 4) illustrates a view of the DSL and how user and microwave instructions are composed into a cooking workflow. An application of UML profiles would permit this general view to be specialized for each specific product and its associated cooking instruction set. However, in line with the principles applied in this work, the primary way of composing such instructions is through an icon based graphical interface.

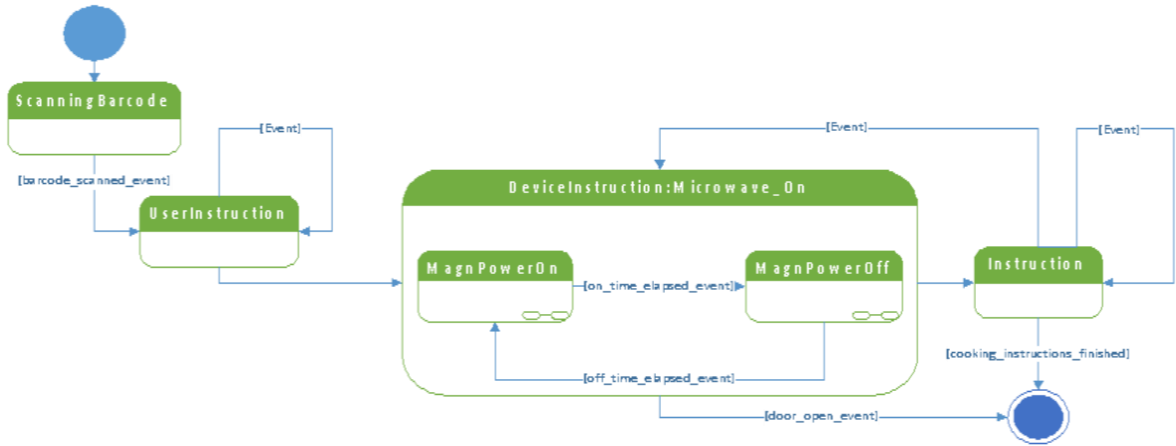

**Figure 4.** Representation of the UML state diagram for the CSA.

### 4. The community supported appliance systems architecture

The creation of a system that is adaptable to users with various needs, a DSL-based system which enables interactive elements to be created "by the users for the users", was one of the main objectives of the project. The prototype CSA was composed of:
1. A cloud-based repository, in which all the information for the user cooking instructions and the microwave cooking instructions can be uploaded and stored in JSON format;
2. A functional and usable representation of the repository, accessible via a customized graphical user interface on a web app or browser for tablet or desktop systems;
3. An add-on module for microwaves [25] composed of a Raspberry PI, with Wi-Fi and Bluetooth modules, a barcode scanner, a touchscreen and a set of speakers;
4. Additional device and environmental sensors such as weight sensor, smoke sensor and door sensor (applied on the microwave's door) were used.

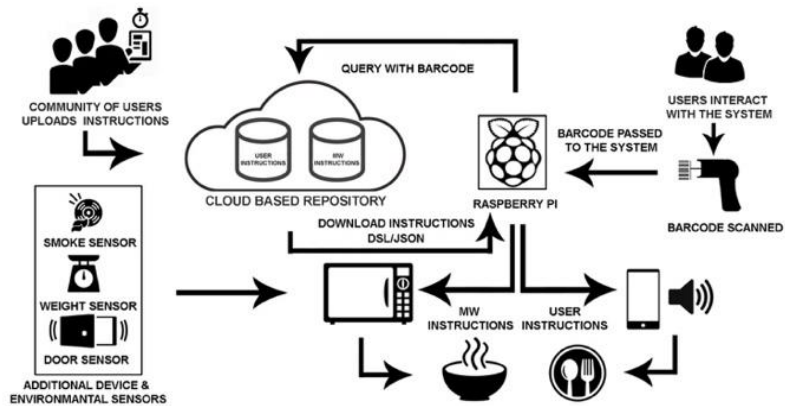

**Figure 5.** The Community Supported Appliance system's architecture.

The interaction process follows five distinct steps as shown in figure 6.

To use (1) a customized graphical user interface, users with disabilities, with the guidance of an occupational therapist or a member of staff can upload specific food information into the system's cloud-based repository. The information required is food name, number of the associated barcode, food category (for easier indexing) and an associated picture/image.

To use (2) the customized graphical user interface so the users and/or members of the care staff will be guided through the process of uploading two types of additional information for the food product: (a) customized user cooking instructions related to which actions the user will perform, with an image, icon or video-clip as a reference, a short description and an associated audio file; (b) processable microwave cooking instructions related to actions the Raspberry Pi will have to perform to control the microwave such as setting the temperature or wattage, cooking time and weight of the food package.

Users (3) will scan the barcode of a food item using the barcode scanner associated with the Raspberry PI module to begin the cooking process.

The system (4) recognizes product barcodes and searches for associated resources in the repository. Each food preparation resource, consisting of a customized user and micro-wave cooking instructions, associated with that barcode, will be downloaded from the cloud repository in real time to the Raspberry PI module which controls the microwave, the touch screen and the speakers.

The workflow (5) of user and device instructions will now be executed for the scanned food item. The user instructions will be played as sequenced video clips on the touchscreen and as audio files on the speakers prompting the users to engage in the cooking process and the device instructions will be used to activate the microwave to begin executing the cooking instructions according to the defined workflow.

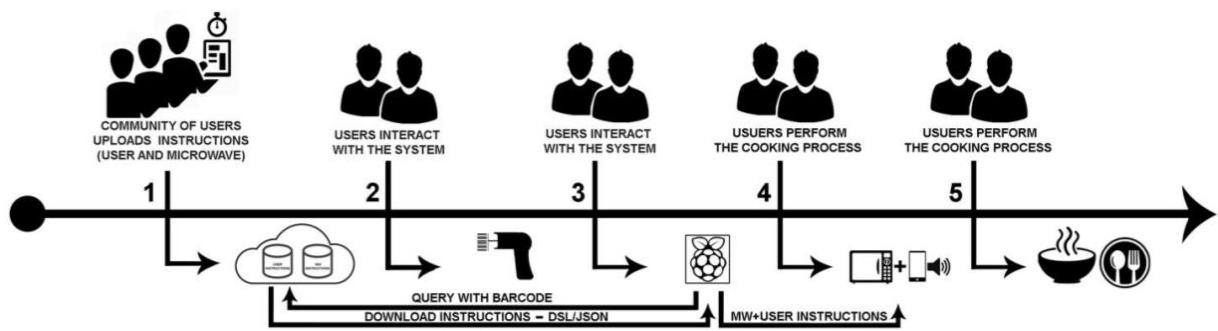

**Figure 6.** The interaction process between the community, the user and the Community Supported Appliance.

According to the co-Design approach undertaken with users, it was possible to define a physical and graphical user interface that supports users in creating personalized user instructions for cooking different types of ready-made foods as well as device instructions for the microwave oven.

A further important design decision that was made was to keep the animations and the icons as simple as possible and not personalized, in order to avoid any confusion that might occur when a service user sees a photo of a real item [10].

Finally, the CSA has the potential to give different types of feedback to the user, before and during the cooking process. Two embedded functions at the core of the interaction system provide this level of feedback: suggestions and alerts.

Suggestions work by providing a set of instructions in order to guide the user in the cooking/heating process. The system will be able to give different personalized instructions depending on the level of the user's abilities.

Alerts address the fact that it is known that the heating/cooking process is associated with safety concerns, such as the risk of fire or burns, the risk of over-heating the meal, the risk of flashings if a metal utensil has been left inside the microwave or simply if the door has been left open. In any such scenarios the system will provide audio and visual alerts to the user to guide them.

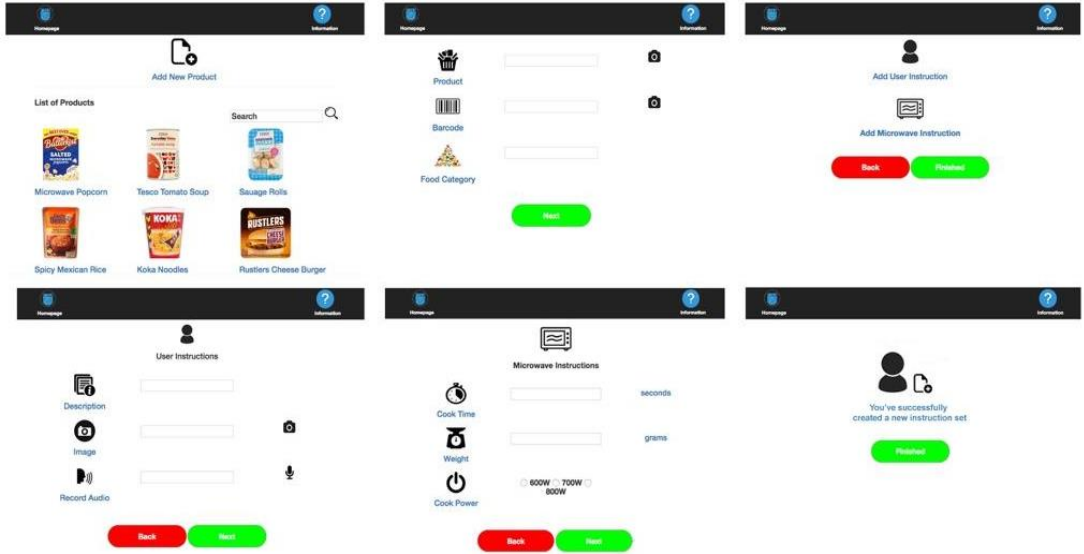

**Figure 7.** HTML-based interface for uploading food information and cooking instructions.

## 5. Conclusion

Software is too often not designed with users and for users. This is particularly true for people with intellectual and physical disabilities. But one of the main research questions was: is there a simple grammar that would allow the translation of the language of the user into the language of the microwave oven?

As a part of a long-term project developed by the team [10], one of the interesting findings that emerged during this research activity was that working both in the hard-ware and software domains, while involving end users in the design process, resulted in a more comprehensive approach to develop the CSA. The CSA, developed with the users, effectively becomes an interactive type of smart object that could be named as "App-cessory", designated by combining the word 'application' and 'accesso-ry'. This can be considered as a new paradigm of UC, where a single physical device that performs a specific function can work only and exclusively with a certain type of application or Web-based service.

The novel aspect of the application of a DSL in the cooking domain for users with special needs, was the co-designed approach used for the whole project. We are conscious that the use of a barcode scanner could have limitations, related to specific foods, but the feedback derived from the first tests with the user group has shown that the microwave oven is mostly used for heating "ready-made" foods [10]. In perspective, the DSL related to the field of "cooking instructions" could be applied not only to different home appliances such as kettles, microwaves, cookers, but also have the potential to be applied within different scenarios, like the home environment, within public spaces or private/public transportation contexts.

**Acknowledgements**

Dr. Matteo Zallio wrote section 1 and 4. Dr. Paula Kelly wrote section 2. Dr. Damon Berry wrote section 3. All authors contributed equally to section 5.  We are grateful for the immense contribution from Mr. Barry Cryan who contributed to research, development and design. This research was done in collaboration with Saint John of God Community, based in Dublin, Ireland. Its staff and community members provided fundamental knowledge and extreme cooperation in ideation, research and codesign stages of the project.